\documentclass[sn-mathphys-num]{sn-jnl}

\usepackage{graphicx}%
\usepackage{multirow}%
\usepackage{amsmath,amssymb,amsfonts}%
\usepackage{amsthm}%
\usepackage{mathrsfs}%
\usepackage[title]{appendix}%
\usepackage{xcolor}%
\usepackage{textcomp}%
\usepackage{manyfoot}%
\usepackage{booktabs}%
\usepackage{algorithm}%
\usepackage{algorithmicx}%
\usepackage{algpseudocode}%
\usepackage{listings}%

\theoremstyle{thmstyleone}%
\theoremstyle{thmstyletwo}%

\theoremstyle{thmstylethree}%

\begin{document}


\title[Structuring the outer solar system]{Trajectory of the stellar flyby that shaped the outer solar system}


\author*[1]{\fnm{Susanne} \sur{Pfalzner}}\email{s.pfalzner@fz-juelich.de}
\author[1]{\fnm{Amith} \sur{Govind}}\email{am.govind@fz-juelich.de}
\author[2]{\fnm{Simon} \sur{Portegies Zwart}}\email{spz@strw.leidenuniv.nl}

\affil*[1]{\orgdiv{J{\"u}lich Supercomputing Centre}, \orgname{Forschungszentrum J{\"u}lich}, \postcode{52428} \city{J{\"u}lich},  \country{Germany}}

\affil[2]{\orgdiv{Leiden Observatory}, \orgname{Leiden University}, \orgaddress{\street{PO Box 9513},\city{Leiden}, \postcode{2300}, \country{the Netherlands}}
}


\abstract{Unlike the Solar System planets, thousands of smaller bodies beyond Neptune orbit the Sun on eccentric ($e >$ 0.1) and ($i>$ 3$^\circ$) orbits. While migration of the giant planets during the early stages of Solar System evolution can induce substantial scattering of trans-Neptunian objects (TNO), this process cannot account for the small number of distant TNOs ($r_p >$ 60 au) outside the planets' reach. The alternative scenario of the close flyby of another star can instead produce all these TNO features simultaneously, but the possible parameter space for such an encounter is vast. Here, we compare observed TNO properties with thousands of flyby simulations to determine the specific properties of a flyby that reproduces all the different dynamical TNO populations, their location and their relative abundance and find that a 0.8$^{+0.1}_{-0.1}$ {\mbox{$M_{\odot}$}} 
star passing at a distance of 
\mbox{$ r_p =$ 110  $\mathbf{\pm}$ 10 au,}  inclined by 
\mbox{$i$ = 70$^\circ$ $^{+5}_{-10}$} 
gives a near-perfect match. This flyby also replicates the retrograde TNO population, which has proved difficult to explain. Such a flyby is reasonably frequent; at least 140 million solar-type stars in the Milky Way are likely to have experienced a similar one. In light of these results, we predict that the upcoming Vera Rubin telescope will reveal that distant and retrograde TNOs are relatively common.}


\maketitle

\section{Introduction}
\label{sec:intro}

The solar system planets accumulated from a disc of dust and gas that once orbited the Sun. Therefore, the planets move close to their common plane on near-circular orbits. About 3000 small objects have been observed to orbit the Sun beyond Neptune ($r_p >$ 35 au); surprisingly, most move on eccentric and inclined orbits \cite[][]{Gladman:2021, Kavalaars:2020}. Therefore, some force must have lifted these trans-Neptunian objects (TNO) from the disc where they formed and altered their orbits dramatically. One popular hypothesis is that the planets originally were in a more compact configuration; the TNOs formed between them and were scattered outwards when the planets moved to their current locations \citep[e.g., ][]{Fernandez:1984, Hahn:1999, Gomez:2003, Morbi:2003, Levison:2008, Raymond:2018, Gladman:2021}. 

However, three distinct TNO dynamical groups are incredibly challenging to explain by the original planet scattering: (i) the cold Kuiper belt objects (KBOs) moving on nearly circular orbits close to the plane, (ii) the Sedna-like TNOs orbiting at large distances \mbox{($r_p>$ 60 au)} on highly eccentric orbits \mbox{($e >$ 0.5)} and (iii) TNOs with high inclination  ($i>$60$ ^\circ$) \mbox{\citep{Brown:2004, Trujillo:2014, Sheppard:2019, Gladman:2009, Chen:2016}}.  While only three Sedna-like objects and two highly inclined TNOs are known so far, they are the make-or-break test for any outer solar system formation theory. Their existence, especially the observed clustering among the Sedna-like and high-inclination TNOs, is unlikely to stem from scattering by the planets \citep{2014MNRAS.444.2808P, Gladman:2021}.

Here, we build on a completely different hypothesis for the TNOs' origin \cite{Kobayashi:2001, Kenyon:2004, Kobayashi:2005}. In this model, the TNOs formed in the outer solar system ($>$ 30 au) and the close passage of another star catapulted them to their current orbits. This hypothesis was initially overlooked as such close flybys were deemed too rare. However, recent ALMA observations reveal that close stellar flybys seem to be relatively common \citep{Dai:2015, Rodriguez:2018, Rosa:2019, Winter:2018b, Akiyama:2019, Menard:2020}. Recently, this scenario has gained renewed interest due to simulations showing that flybys can produce a cold Kuiper belt population and Sedna-like objects \cite{2014MNRAS.444.2808P, Pfalzner:2018, Moore:2020}. These proof-of-principle studies considerably strengthened the flyby hypothesis. However, the possible flyby parameter space has remained relatively large, and the resulting predictions remained vague. More precise predictions are essential to decide between the competing hypotheses. Here, we present the essential next step -- we provide the close-to-exact parameters of the potential outer solar system shaping fly. The resulting predictions are distinct and testable by the $\approx$ 40 000 TNOs awaiting discovery when the LSST becomes operational \citep{LSST_book}. The TNOs orbiting in the opposite direction as the planets ($i >$ 90$ ^\circ$) -- so-called retrograde TNOs -- may be the key to this decision.

\section{Results}
\label{sec:results }

Our exhaustive numerical parameter study consists of over 3000 individual simulations modelling the effect of a stellar flyby on a planetesimal disc surrounding the Sun extending to \mbox{150 au} and \mbox{300 au, respectively}.  Such sizes have been observed to be typical for protoplanetary and debris discs \citep{Andrews_2020,Hendler_2020}. We vary the mass of the perturber, $M_p$, its perihelion distance, $ r_p$, and the relative orientation of its path in terms of inclination, $i$, and angle of periastron, $\omega$, and the size of the disc, $R_d$. 

We systematically rejected any flyby that failed to \emph{quantitatively} match the observed TNO population. {This means that any successful candidates had to reproduce the location in the $a, e, i$ parameter space and the relative population sizes of the cold KBOs and the Sedna-like objects. The latter are particularly important as, unlike the resonant TNOs, their relative numbers and orbits are largely unaffected by interactions with Neptune after the flyby, expressed by the Tisserand parameter $T<$3.05. In addition, we demanded that the planet orbits remain unperturbed (for details, see Methods section). Only three flybys met our strict criteria for an excellent quantitative match to the observed TNOs (see Table 1). These three flybys produced the hot, cold, and Sedna-like TNOs in the observed relative quantities and in the right places in the multi-dimensional parameter space.  Each best-fit model emphasised different TNO dynamical groups in the selection process. Still, their parameters are so similar that one can combine them into a single flyby scenario with a remarkably small error bar.

For a {\it parabolic} flyby, we find that a star with mass $M_p =$ 0.8$^{+0.1}_{-0.1}${\mbox{$M_{\odot}$}}  at a perihelion distance \mbox{$ r_p =$ 110  $\mathbf{\pm}$ 10 au} inclined by \mbox{ $i$ = 70$ ^\circ$ $^{+5}_{-10}$} and an angle of periastron of \mbox{60$ ^\circ$-- 90$ ^\circ$} provides the best candidate for an outer solar-system-shaping flyby based on current data. The spatial orientation is given relative to the plane of the pre-flyby disc. For an illustration of the flyby dynamics, 
see \mbox{Fig. 1} and the Supplementary video. The past flyby orbital parameters are shown by Fig. 2, left. We performed higher-resolution simulations for models A -- C with 10$^5$ tracer particles and modelled two disc sizes (150 au and \mbox{300 au} - models A1 and A2, respectively).
Interestingly, these parameters are fairly consistent with those of another flyby scenario \cite{2015MNRAS.453.3157J}, which argues that a $1.8$\,M$_\odot$ star would have passed the Solar system at $ r_p =$ 227 au inclined by 17$^\circ$ -- 34$^\circ$. The different mass can be explained by the type of encounter studied: where \cite{2015MNRAS.453.3157J} adopted an exchange interaction to abduct Sedna from the intruder, whereas here we argue that Sedna (and the other KBOs) are native to the Solar system.

The flyby probably happened several Gyr in the past; thus, how much change the orbital parameters on such a time scale? Investigating the long-term evolution of the TNO population is computationally expensive. Therefore, we studied only the period of 1 Gyr after the flyby. The general outcome remains very similar (see Fig. 2, middle). The changes include an increase in low-inclination TNOs, improving the match to the cold TNO population and filling in the low-inclination distant TNOs missing immediately after the flyby. Thus, the long-term evolution leads to an even better fit.

The final model delivered a surprise: The best-fit flyby created retrograde TNOs despite them \emph{not} being part of the selection process. So far two retrograde TNOs have been confirmed – \mbox{2008 KV$_{42}$} and \mbox{2011 KT$_{19}$}, both having relatively small periastron distances \mbox{($r_p <$ 30 au,} \mbox{$a>$ 30 au)} and are inclined by 103.41$ ^\circ$ and \mbox{2011 KT$_{19}$} by 110.15$ ^\circ$, respectively. 
An additional TNO is suspected of moving on a retrograde orbit -- 2019 EE$_6$ -- but its orbit is currently not well constrained. It is more distant ($r_p >$ 30 au) and may be closer to the plane \mbox{($>$ 150$ ^\circ$).}

Eventually, high-inclination TNOs could be crucial when deciding between different hypotheses. Retrograde TNOs, as such, provide a challenge for the planet instability model. Adding a distant planet (Planet Nine) appeared to solve the problem \citep{Batygin:2016,Batygin:2024}. This combined model can account for retrograde TNOs  with \mbox{$r_p<$ 30 au} and \mbox{$i<$ 150$ ^\circ$} like \mbox{2008 KV$_{42}$} and \mbox{2011 KT$_{19}$} \cite[\mbox{see Fig. 1} in ][]{Batygin:2016}.
However, distant, highly inclined TNOs ($r_p>$ 30 au, $i>$150$ ^\circ$), if existing, may provide a challenge also for the planet nine model. 

Conversely, retrograde TNOs might also be the key to determining the primordial size of the solar system disc. The maximum inclination of retrograde TNOs is directly related to the primordial disc size (see Fig. 3). The inclinations of 2008 KV$_{42}$ and \mbox{2011 KT$_{19}$} (103.41$ ^\circ$ and 110.15$ ^\circ$) demand that the Sun's primordial debris disc must have extended to at least 
\mbox{$R_{d} \ge $ 65 au.} Close to the plane retrograde TNOs would argue for an even larger size \mbox{($R_{d} \ge $150 au).  } Using this relation, retrograde TNOs detected in the future will enable setting stringent bounds on the primordial disc size.

Currently, only the nearest and brightest TNOs are observable, and high-inclination and very eccentric objects are challenging to detect. The right panel of Fig. 2 supplies a sneak preview of the TNO discoveries we expect from the here presented flyby scenario. It shows that the clustering among the known highly inclined TNOs \cite{Chen:2016} and Sedna-like objects is part of a much larger pattern caused by the flyby. It will be interesting to see this pattern emerge when more TNOs are discovered. Although the pattern becomes slightly less distinctive on Gyr timescales due to secular effects (see middle panel), the clustering as such persists (see Fig. 2, middle).

The information about the flyby parameters enables us to predict how the relative sizes of different TNOs dynamical groups will change when the observable space expands (see Supplementary Figure 1 and Supplemantary Table 1). Matching the observations, Sedna-like TNOs make up only about 0.1\% of all TNOs in model A--C in the currently observationally accessible space. However, this will increase to 7\% by the end of the ten-year observation campaign of the Vera Rubin telescope as more distant TNOs become observable. Likewise, we anticipate an increase in the fraction of retrograde TNOs from 0.15\% to about 5\% as the discovery space expands. Although some of the expected retrograde TNOs may orbit close to the plane, we foresee most of them moving at high inclinations from the plane.

However, we caution against overinterpreting Fig. 2. To some extent, we expect the non-detection of TNOs in covered areas. Neither the size nor the structure of the primordial solar disc is known. Any change -- smaller size or ring structures -- in the primordial disc leads to "holes" in the parameter space indicated in Fig. 2. Indeed, such gaps could even help to determine the solar disc's structure before the flyby. Conversely, if TNOs are found in areas not predicted by Fig. 2 even after including the long-term evolution, this would challenge the presented hypothesis. However, its falsifiability makes the flyby hypothesis methodologically so strong.

So far, we have concentrated on the bound TNO population beyond \mbox{30 au.} However, while leaving the planetary orbits undisturbed, the flyby injects many TNOs  ($\approx$9\%  of the initial disc mass $m_i$) into the area inside 30 au. These injected TNOs move on high eccentricity ($e>$ 0.4), high-inclined \mbox{($>$ 30$^\circ$)} regularly revisiting the trans-Neptunian region \mbox{(60 au – 200 au)}. At the same time, a considerable fraction (26\%) of the TNOs become unbound from the Sun (see Supplementary Figure 2), and the perturber captures 8.3\% of the material initially bound to the Sun (model A1). While moving on highly eccentric orbits, some of these captured solar TNOs come incredibly close to the perturber star \mbox{($r_p^{min}$ = 0.73 au)}. These TNOs move well within the ice lines of this system, where volatiles evaporate.

\section{Discussion}

The known TNO population is subject to many different biases \citep[for a discussion see, ][]{Bannister:2018,Bernardinelli:2022}, and 
likely represent only a fraction ($<$1\% -- 10\%) of the entire population. New TNOs are constantly discovered, some with entirely unexpected orbital properties \cite{Shephard:2016,Chen:2016}. Thus, searching for flyby parameters best fitting the observations presented here can only be a first step. Once a significant portion of the TNOs is known, this procedure must be repeated, and the flyby parameters adjusted accordingly. Despite these reservations, we expect the final best-fit parameters to be similar. After all, the model must still account for the Kuiper belt, Sedna-like and retrograde TNOs at the currently known positions in the multi-dimensional parameter space. Alternative hybrid schemes combining planet scattering with one or more flybys have been suggested \citep{Nesvorny:2023}. However, it is an open question whether such hybrid scenarios have predictive power.

When would this flyby have occurred? Close encounters are most frequent during the first 10 Myr of a star's life when it is still part of its birth cluster. Simulations find that typically \mbox{8\% -- 15\%} of all solar-type stars experience an encounter reducing the unperturbed area to 30 au - 50 au in favourable environments (similar to NGC 2244 and M44) \cite{Pfalzner:2020}. Even in low-density clusters, $\approx$ 1\% of solar-type stars experience such an encounter. Putting this number in perspective:  in the first 10 Myr of their life, at least 140 million solar-type stars (possibly ten times more) have experienced an encounter similar to the Sun's in the Milky Way. In $\approx$10\% of these cases, the encounter was with a similar mass perturber  \mbox{$M_p$= 0.6 -- 1.0 {\mbox{$M_{\odot}$}}} at approximately the same periastron distance ($r_p$ = 90 au -- 130 au) as the Sun's flyby. Close flybys became less frequent after the solar birth cluster expanded and dissolved at the end of the star formation process. However, the 4.55 Gyr that passed since the solar system formed more than outbalances the much lower encounter frequency, summing up to a probability of 20\%-- 30\% likelihood for a late encounter \cite{Pfalzner:2018}.  However, due to the stellar velocity dispersion increasing with the Sun's age, these flybys would be mainly on highly hyperbolic orbits. Hyperbolic flybys are much less efficient in exciting the orbits of TNOs. Therefore, the question of whether a later flyby could lead to a similarly good match require further study. 

The flyby scenario neither excludes the planets forming in a more compact configuration nor the existence of a primordial Oort cloud. Planet migration could have scattered additional objects into the trans-Neptunian region, contributing to the hot Kuiper belt population without altering the Sedna-like or retrograde TNO populations. Even if the planets were located at their current positions at the time of the flyby, they would have been unaffected by the flyby except Neptune. If Neptune were at its current distance at the time of the flyby, it would have been shielded from the effect of the flyby in 25\% of case, staying in the gravitational shadow of the perturber – meaning flying behind the perturber star  (see Supplementary Figure 2). 

If the Oort cloud existed before the flyby, it would have been severely affected, but \emph{not erased}. A flyby of the given parameters would have left a sufficiently large number of TNOs \mbox{($\approx$ 15 \%)}  bound to the Sun to account for the current estimates of the Oort cloud mass. Besides, the Oort cloud might have been simultaneously enriched by TNOs with $a \gg$ 10$^4$ au, originally belonging to the outer disc \mbox{(80 au $<r_p < R_d$)} and planetesimals initially being part of the potentially existing perturber Oort cloud \cite{2021A&A...647A.136P}.

Finally, one may speculate whether the probability of the perturber's planetary system developing life increased by the flyby. The probability would have been higher if the flyby happened not during the first 10 Myr but later when pre-forms of life had already developed.

\section{Conclusion}

We demonstrated that the flyby of star of mass \mbox{$M_p =$ 0.8$^{+0.1}_{-0.1}${\mbox{$M_{\odot}$}}} 
passing on a parabolic orbit at a perihelion of
\mbox{$ r_p =$ 110  $\mathbf{\pm}$ 10 au,} 
at an inclination of \mbox{$i$ = 70$^\circ$ $^{+5}_{-10}$} explains several unaccounted-for outer solar system features. It quantitatively reproduces the orbital properties of the cold Kuiper belt population, Sedna-like objects and high-inclination TNOs. Unexpectedly, this flyby also accounts for the otherwise difficult-to-explain retrograde population.  The model's beauty lies in its simplicity and ability to make specific predictions. These predictions include a distinct clustering in $a$-, $e$-, $i$-space and a rise in the relative fraction of retrograde and Sedna-like TNOs. Future TNO discoveries may reveal the primordial solar system disc's size and structure.

\section{Method}

\subsection{Flyby simulations and selection procedure}

Our parameter study consists of 3080 individual simulations
modelling the effect of stellar flybys on a planetesimal or debris disc surrounding the Sun. The aim was to find the subset of simulations that produce the various dynamics groups in the observed quantities and locations in the multi-dimensional parameter space. Previous work \citep{Pfalzner:2018} found that the most promising parameter space for finding the most challenging TNO dynamical groups - entails perturber masses in the range \mbox{$M_p= $ 0.3 -- 1.0 {\mbox{$M_{\odot}$}}}, periastron distance  \mbox{$\mathrm{r}_{\mathrm{peri}}$ = 50 -- 150 au}, inclinations $ i $= 50$^{\circ}$ -- 70$^{\circ}$, and angles of periastron \mbox{$\omega$  = 60$^{\circ}$ -- 120$^{\circ}$}. We scanned this parameter space in mass steps of  0.1 {\mbox{$M_{\odot}$}}, $r_p$ in steps of 10 au, $i$ in units of 5$^\circ$and the \textcolor{teal}{$\omega$} with a variation of 10$^\circ$.

The simulations start with an idealised thin disc \cite{Pringle:1981} represented by $N$=10$^4$ mass-less tracer particles. Taking the observed sizes of typically 100 -- 500 au of protoplanetary and debris discs for guidance \citep{Andrews_2020,Hendler_2020}, we model disc sizes of \mbox{150 au} and \mbox{300 au.} We treat model the $N$ gravitational three-body interactions between the Sun, the perturber star and each of the $N$ test particles \citep{Kobayashi:2001,Musielek:2014,Pfalzner:2018}. Self-gravity and viscosity effects are negligible, as the interaction time is short ($<$ 4000 yr) and the disc's mass is considerably smaller than the Sun's ($m_d \ll$ 0.001 $M_{\odot}$).  We use a Runge-Kutta Cash-Karp scheme to determine the particle trajectories.  The simulations start and end when the force of the perturber star on each particle is less than 0.1\% 
\cite{Breslau:2014}. We optimise the computational effort by using an initial constant particle surface density to obtain a high resolution in the outer parts of the disc. We then post-process the data by assigning different masses to the particles to model the actual mass density distribution \citep{Hall:1996,Steinhausen:2012}.

 We set strict standards for matching observations with simulations, rejecting 99.9\% of all simulated cases. Nevertheless, this computational expense paid off. We obtained a near-perfect match to the known TNO population. We tested only for those TNOs not strongly coupled to Neptune  (T$_\mathrm{N} >$ 3.05, where $T$ is the Tisserand parameter). Thus, most resonant TNOs were excluded from the comparison.  Similarly, we did not consider TNOs with $a >$10,000 au as more distant encounters and the galactic potential could affect their orbits over Gyr timescales. 
After the flyby, some objects enter into a resonant orbit with Neptune during our long-term simulation. They are not visible in Fig. 1 since they do not meet the $T_N >$ 3.05 threshold. Likely the number of resonant objects is small because the simulation only covers the first 1 Gyr,  additional resonant TNOs may be produced over extended periods. A higher resolution of the disc population would also required to describe  this process adequately. Besides, resonant TNOs may be produced if Neptune migrated outward after the flyby.

We used a decision tree-based inspection method, first selecting the flybys that avoid strong perturbations inside 30 au -- 35 au. We used the  approximation,  
$ r_{\text {d}}=0.28 \times M_{\text {p}}^{-0.32} r_{\text {peri}}$
\cite{Breslau:2014}, as an indicator of the radial distance r$_{\text {d}}$ up to which the disc remains largely undisturbed. This equation applies only to coplanar encounters while we study inclined encounters. Therefore, we slightly extend the parameter space to account for the difference. A subset of 490 simulations fulfilled the criterion of an unperturbed population up to \mbox{30 au -- 35 au.} Here, we assume that the planets orbit at their current locations. If the solar system was in a more compact configuration during the flyby, the constraints would relax. Next, we retained only flybys that produce a cold Kuiper belt population and Sedna-like objects in the suitable regions of the parameter space. Only a small subset clustering around perturber masses 0.7 - \mbox{0.9 {\mbox{$M_{\odot}$}}} and periastron distances of 90 au -- 110 au fulfils this criterion. Among the few remaining possibilities, additional cases can be excluded on more stringent criteria. For example, among the $r_p$=110 cases, higher-mass perturbers \mbox{($M_p \ge$ 0.9 {\mbox{$M_{\odot}$}})} tend to produce too few cold Kuiper belt objects, while lower-mass perturbers ($M_p \le$ 0.7 {\mbox{$M_{\odot}$}}) have difficulties reproducing the high eccentricity TNOs. For 0.8 {\mbox{$M_{\odot}$}} perturbers, only perihelion distances of 100 au and \mbox{110 au} can produce the right size of the unperturbed region. The closer encounter (100 au) produces 80\% fewer cold TNOs than the 110 au perturber. Hence, a 0.8 {\mbox{$M_{\odot}$}} perturber passing at a periastron distance of 110 au best fits the observational data.

We simultaneously tested for the inclinations and the argument of perihelion of the perturber's orbit. Again, the relative number of cold Kuiper belt objects is a key element. Supplementary Figure 3 shows the dependence of the number of cold population particles as a function of $i$ and $\omega$ for a flyby with $M_p =$ 0.8 {\mbox{$M_{\odot}$}} at $ r_p =$ 110 au. The cold population decreases significantly below 70$^{\circ}$  inclination and 80$^{\circ}$  argument of perihelion. While above these values, the simulations do not reproduce the inclination and eccentricity distributions of the TNOs correctly. Hence, an inclination of 70$^{\circ}$  and an angle of perihelion of 80$^{\circ}$  produce the best fit.

We tested the best-fit flyby to check their influence on the giant planets orbits. Our criterion is that the changes in $i$ and $e$ due to the flyby should be less than the difference of currently observed planetary orbits from being circular and in the plane. Neptune's orbit is more vulnerable than those of the other planets. However, the key parameter is the orbital position at the moment of flyby. Even Neptune's orbit remains nearly unaffected  ($\Delta <$ today's $e$ and $i$) at the locations indicated in blue in Supplementary Figure 2). Uranus's eccentricity remains unaffected; however, small ranges of positions are excluded because the inclination is slightly higher (1$^o$) than today's (0.7$^o$). The influence on Jupiter and Saturn is negligible, independent of orbital location.

When performing such a comparison, one faces two challenges: (i) the biases in the known TNO population \cite{Kavalaars:2020, Gladman:2021} and (ii) the fact that the size of the primordial disc is unknown. Therefore, we determined three best fits emphasising different populations (see Table 1). Model B gives a slightly larger cold population than A 1 (see Supplementary Figure 4). Model C produces more high-inclination objects (see Supplementary Figure 6). Models A1 and A2 only differ in their assumed disc sizes of 150 au and 300 au, respectively. 

While this method was labour-intensive, it was the most reliable approach. Automated statistical methods \cite{Jilkova:2015, Moore:2020} generally test against deviations from the median or mean. We find that taking a mean as the decision basis fails to account for multiple clustering in TNO dynamical groups, especially in multidimensional parameter space. Besides, various observational biases make it problematic to compare ``unbiased" simulation results in an automated way. 
We also tested using the observation simulator OSSOS \citep{Bannister:2018}, applying the default absolute magnitude distribution to assign magnitudes to the test particles. We find that for model A1, 70 objects of our simulated objects should be currently observable. However, rating the quality of this match in an automated way faces the problem that the result depends sensitively on the size of the chosen comparison parameter space.

\subsection{Long-term evolution}

Determining the long-term evolution after the flyby requires a high-precision integrator, which makes these simulations computationally expensive. Therefore, we modelled the long-term only for a subset of the results consisting of all particles  fulfiling the conditions: 
\mbox{35 au $< r_p <$ 90 au,} \mbox{$i<$ 60$^\circ$} and $a <$  2000 au.  These TNOs correspond to $\approx$20 \% of the total TNO population and roughly to the TNOs that should be visible with instruments like the Vera Rubin telescope. In addition to the test particles from the flyby simulation, the four outer giant planets were included in the long-term simulation. We start with the particle positions and velocities at 12 000 years after the perihelion passage. Using the GENGA code \cite{Grimm_2014}, we follow the trajectories of the test particles for the consecutive 1 Gyr. These trajectories are determined using a hybrid symplectic integrator.

\subsection{Flyby frequency determination}

We determined the occurrence rate of such close flybys in different cluster environments ranging from short-lived low-N clusters to massive, compact, long-lived clusters. We performed an extensive set of \mbox{N-body} simulations using the code Nbody 6++ \cite{Aarseth:2003}. In these simulations \cite[for details of the simulations, see][]{Pfalzner:2020},  the cluster development matches that of observed clusters regarding the temporal development of the cluster mass and size with cluster age. We computed hundreds of realisations so that the results have high statistical relevance. We record the parameters of any close interaction between two stars and use this information in a post-processing step to determine the effect of each encounter on the disc size (equalling the unperturbed area after an encounter). We study the sub-set of solar-type stars and test for the frequency of encounters leading to a 30 -- 50 au-sized unperturbed disc.  We also test for solar-type stars encountering a perturber of mass \mbox{0.6 -- 1.0 {\mbox{$M_{\odot}$}}  at a distance of 90 - 130 au, similar to our best-fit results.}

\subsection{Toy model for effect on the Oort cloud}

We estimated the effect of such a flyby on a potentially existing Oort cloud. Our toy model consisted of 10 000 particles randomly distributed in a 100 000 au-sized sphere surrounding the Sun. We simulated model A's flyby effect on this Oort cloud. The particle trajectories are calculated using the REBOUND N-body code \citep{rebound} employing IAS15, a 15th order Gauss-Radau integrator \citep{reboundias15}. 

\newpage

\bmhead{Data availability} The data of the complete parameter study are available on the DESTINY database under the following link https://destiny.fz-juelich.de. 

\bmhead{Code availability} The codes REBOUND and GENGA are open access codes. The DESTINY code will be available upon reasonable request. However, the DESTINY database (https://destiny.fz-juelich.de) also allows to perform diagnostics online. It allows to reproduce Fig. 2, and also similar plots for the entire parameter study. A complete illustration of the dynamics of the flyby scenario of model A1 is available in the Supplementary video.

\bmhead{Author contributions}
Conceptualization, S.P.; Simulation of the parameter study, long-term evolution and effect of Oort cloud: A.G.; Diagnostic: S.P., A.G.; Comparison to observational data: A.G., S.P., S.P.Z.; Writing –  S.P., A.G., S.P.Z.; Funding Acquisition \& Resources, S.P., Supervision, S.P.; Data Curation, A.G.

\bmhead{Acknowledgments}

We want to thank Sonja Habbinga for her assistance in the visualization of model A1,  M. Bannister and R. Dorsey for advising us on interpreting TNO survey results and the OSSOS simulator and F. Wagner for supporting us in implementing the code GENGA on the FZJ system. SP has received funding for this project through the grant  450107816 of the Deutsche Forschungsgemeinschaft. 

\bmhead{Competing interests}
The authors declare no competing interests related to the topic of this paper.

\newpage

\begin{table}[h]
\caption{{\bf Flyby scenarios reproducing the known TNO population}. The first column gives the scenario identifier, column 2  the TNO emphasised sub-group when determining the best fit, column 3 the perturber mass, $M_p$, column 4 the periastron distance of the flyby, $r_p$, column 5 the inclination, $i$, column 6 the angle of periastron, $\omega$, and column 7 the assumed pre-flyby disc size}.
\label{tab:best_fits}
\centering
      \begin{tabular}{llccccrr}
\hline 
\hline
    & emphasis on & $M_p$ [M$_{sun}$] & $r_p$ [au] &  $i$ [$^\circ$] &  $ \omega$ & [$^\circ$]
    & $R_d$ [au]\\   
\hline
A   &  Sedna-like        &  0.8    & 110  & 70   &  80  &   150 -- 300   \\
B   &  Cold Kuiper belt   &  0.8    & 110  & 70   &  90  &   150     \\
C   &  ETNOs                          &  0.8    & 110  & 65   &  60  &   150     \\ 
           \hline    
      \end{tabular}
\end{table}


\noindent

\begin{figure*}[h]
\centering\includegraphics[width=\textwidth]{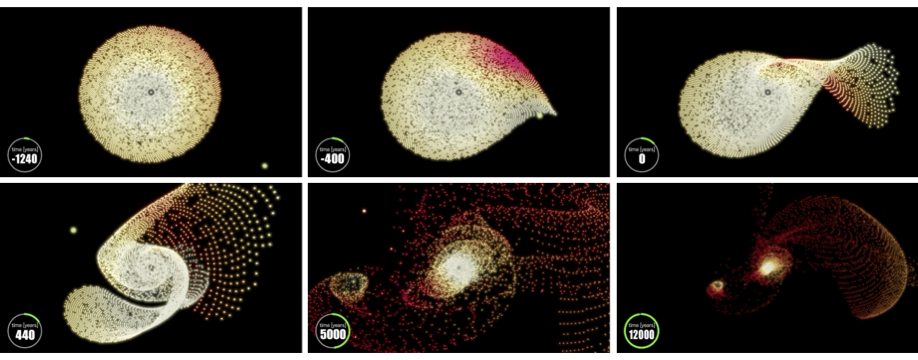}
\caption{{Simulation snapshots of model A.} The perturber star approaches from the bottom right. The sequence shows the typical appearance of two spiral arms, the loss of matter that becomes unbound and the capture of some material by the perturber star. The time is given in years relative to the time of periastron passage. For the first four snapshots,
the size of the real area is kept constant; the last two plots show a zoom-out. The colours indicate the velocities of the test particles relative to the Sun. The complete dynamics is illustrated in the Supplementary video. }
\end{figure*}

\begin{figure*}[h]
\centering\includegraphics[width=\textwidth]{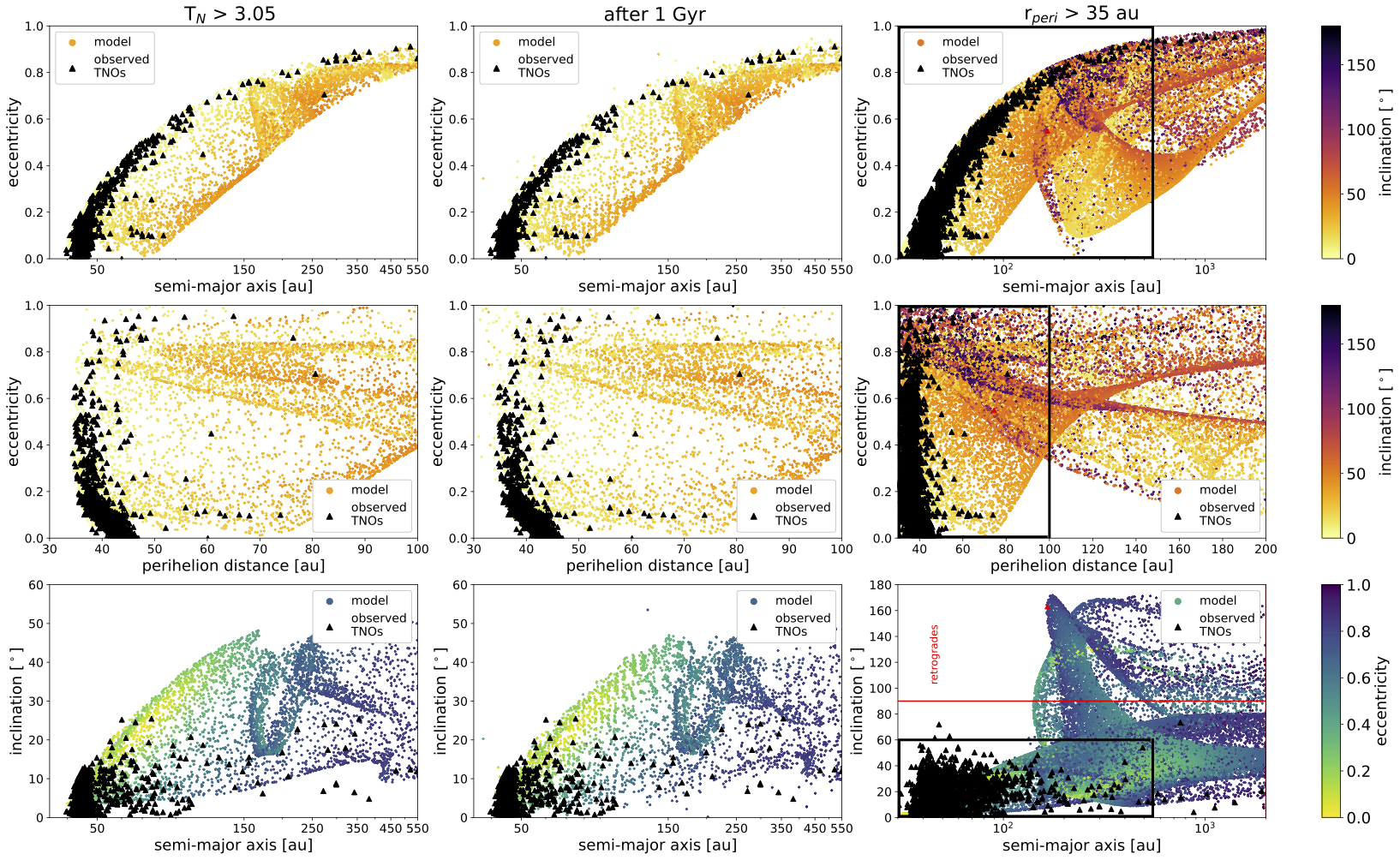}
\caption{{\bf Comparison of TNO orbital parameters between observations and simulation (model A2)}. The coloured symbols show the simulation result, and the black triangles depict the observed TNOs. In the left and middle panel, only the subset of the resulting population \mbox{(35 au $< r_p <$ 100 au,} \mbox{$a <$ 2000 au}, $i <$ 60$ ^\circ$) is shown roughly corresponding the currently observational accessible area. Here only objects with a $T>$ 3.05 were chosen for the comparison. The left and middle panel show the situation 1200 years and \mbox{1 Gyr} after the periastron passage. The right panel provides a map of the predicted properties of the expected TNO discoveries. There, the red triangles indicate the nominal positions of the recently discovered retrograde TNOs. See Supplementary Figures 1--3 in online open access version.}
\label{fig:ETNOs_sednas}
\end{figure*}

\begin{figure}[t]
\centering\includegraphics[width=0.70\textwidth]{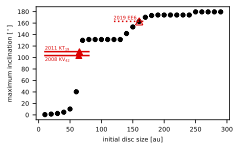}
\caption{{\bf Disc size vs inclination correlation.} Dependence of the maximum inclination on primordial disc size. The filled red triangles show the properties of the confirmed retrograde TNOs, the open symbols of those not yet confirmed. The uncertainty of the orbital parameters of the latter is relatively large, indicated by the indicated red area.}  
\label{fig:scatter_primordial_disc}
\end{figure}

\pagebreak{4}

\end{document}